\documentclass[openacc]{rstransa}

\usepackage[T1]{fontenc}
\usepackage[latin9]{inputenc}
\usepackage{babel}
\usepackage{nccmath}
\usepackage{orcidlink}
\usepackage{placeins}
\usepackage{hyperref}

\usepackage[T1]{fontenc}
\usepackage[latin9]{inputenc}
\usepackage{babel}
\usepackage{nccmath}
\usepackage{orcidlink}
\usepackage{placeins}

\usepackage{graphicx,amsmath,amssymb,color}
\usepackage[normalem]{ulem}
\usepackage{amsfonts}
\usepackage[toc,page]{appendix}
\usepackage{url} 
\usepackage{latexsym}
\usepackage{amsfonts}
\usepackage{algpseudocode}
\usepackage{amsthm}
\usepackage{mathrsfs}
\usepackage{natbib}
\usepackage{color,verbatim}
\usepackage{psfrag}
\usepackage{cleveref}

\renewcommand{\cref}{Fig.~\ref}
\def\beq{\begin{equation}}
\def\eeq{\end{equation}}
\def\bali{{\begin{align}}}
\def\eali{{\end{align}}}

\def\ODM{\Omega_{\rm DM}}
\def\com{\mbox{,}}

\titlehead{Research}

\begin{document}

\title{Symbolic Regression and Differentiable Fits in Beyond the Standard Model Physics}

\author{Shehu AbdusSalam$^{(1)}$\orcidlink{0000-0001-8848-3462},~  
Steven Abel$^{(2)}$\orcidlink{0000-0003-1213-907X},~
Deaglan Bartlett$^{(3,4)}$\orcidlink{0000-0001-9426-7723} and Miguel Crispim Rom\~ao$^{(2)}$\orcidlink{0000-0003-4539-6283}}

\address{$^{(1)}$Department of Physics, Shahid Beheshti University, Tehran, Iran\\
$^{(2)}$Institute for Particle Physics Phenomenology, Durham University, Durham DH1 3LE, U.K.\\
$^{(3)}$ CNRS and Sorbonne Universit\'e, Institut d'Astrophysique de Paris (IAP), F-75014 Paris, France\\
$^{(4)}$ Astrophysics, University of Oxford, Denys Wilkinson Building, Keble Road, Oxford OX1 3RH, UK}

\subject{High Energy Physics, Computational Physics}


\corres{
\email{s.a.abel@durham.ac.uk}}

\begin{abstract}
We demonstrate the efficacy of symbolic regression (SR) to probe models of particle physics Beyond the Standard Model (BSM), by considering the so-called Constrained Minimal Supersymmetric Standard Model (CMSSM). Like many incarnations of BSM physics this model has a number (four) of arbitrary parameters, which determine the experimental signals, and cosmological observables such as the dark matter relic density. We show that analysis of the phenomenology can be greatly accelerated by using symbolic expressions derived for the observables in terms of the input parameters. Here we focus on the Higgs mass,  the cold dark matter relic density, and the contribution to the anomalous magnetic moment of the muon. 
We find that SR can produce  remarkably accurate expressions. Using them we make global fits to derive the posterior probability densities of the CMSSM input parameters which are in good agreement  with those performed using conventional methods. Moreover, we demonstrate a major advantage of SR which is the ability to make fits using differentiable methods rather than sampling methods. We also compare the method with neural network (NN) regression. SR produces more globally robust results, while NNs require data that is focussed on the promising regions in order to be equally performant.    \\
\end{abstract}

\maketitle
\section{Introduction} 

A fundamental challenge in exploring theories Beyond the Standard Model (BSM) lies in confronting their predictions with experimental data. Traditional methods are well-established: we define a parameter space guided by physical reasoning and pragmatic constraints, then compute the low-energy spectra for each point. Initial filters ensure phenomenological viability, for example insisting on the stability of the lightest supersymmetric particle (LSP) in supersymmetric (SUSY) models, and then further calculation determines observables such as the dark matter relic density or the muon anomalous magnetic moment. To analyse the entire parameter space relatively efficiently (or indeed at all), sampling algorithms such as Markov Chain Monte Carlo (MCMC) or nested sampling methods~\cite{Skilling:2004pqw, Skilling:2006gxv, Ashton:2022grj} (e.g., MultiNest~\cite{Feroz:2007kg} and Dynesty~\cite{2020MNRAS.493.3132S}) can then be employed to derive posterior distributions for the model parameters.

While physically transparent, this approach faces a significant computational bottleneck: calculating each observable for a given set of input parameters involves complex, multi-step physical analyses, often encompassing subtle effects such as multi-loop corrections, co-annihilation, and threshold effects. Let us illustrate by describing the chain of computation that yields the dark matter relic density in the Constrained Minimal Supersymmetric Standard Model (CMSSM). In any BSM model we would like to determine how this observable depends on  the fundamental parameters that are mostly set at a very high energy scale close to the scale of quantum gravity. In the case of the CMSSM  there are actually three high scale parameters that we are free to choose, namely the gaugino masses, scalar masses, and universal trilinear coupling. (These are typically set at the Grand Unified Theory (GUT) scale of $2\times 10^{16}$\,GeV). The fourth free parameter is set at the electroweak scale, and is the ratio of Higgs Vacuum Expectation Values (VEVs): these four parameters are denoted respectively as $m_{1/2},~m_0, ~A_0$ and $\tan\beta $.  The relic density  depends on the low scale (by which we mean electroweak scale) values of the couplings and mass spectrum, which in turn depend on the low scale values of all the parameters in the theory. The first step is therefore to find these  low scale parameters  by  evolving the renormalisation group (RG) equations, which are a coupled set of non-linear differential equations which determine how all the parameters of the theory depend on the energy scale. Once these low scale values of the parameters have been obtained, one can then determine the mass eigenstates and eigenvalues, single out the lightest state (i.e. the LSP) which is our dark matter candidate, and then evaluate the scattering cross-section as a function of temperature. Then this information gets fed into the assumed cosmological evolution in order to determine the density at which the LSP would have stopped annihilating in the early universe (the so-called freeze-out density). This involves solving yet another non-linear differential equation, which finally yields the desired  relic density.  If one simply needs to know whether a particular choice of parameters gives a viable relic density or not, this is a great deal of effort! 

Clearly closed form expressions for observables such as the relic density in terms of the model input parameters would be an invaluable short-cut to all of the labour outlined above. Although, in principle, closed form expressions linking the input parameters directly to the observables could be derived analytically, their complexity would typically render them impractical except in simplified or asymptotic regimes. Let us stay with the dark matter relic density for the moment, to appreciate just how difficult this would be in practice. What would it take for some omniscient being to derive symbolic expressions for the relic density using physical analysis? First they would have to determine closed form expressions for the low scale theory that can approximate all of the RG running. This  is itself a very difficult task. At one-loop order, closed form expressions for the running parameters of the theory  {\it can} be obtained when one of the CMSSM parameters, namely $\tan\beta $, is small because there is some quasi-fixed point behaviour in this region of parameter space. Alas, this possibility is mostly excluded experimentally, and moreover typically RG running must be performed to at least two-loop order to achieve sufficient accuracy. Nevertheless, let us suppose it were possible. Next one would have to derive analytic expressions for the mass spectrum from the low energy parameters. For say the scalar superpartners this would amount to being able to find closed form expressions for the eigenvalues of $6\times 6$ matrices. Next one would need to use these expressions along with the couplings to determine the LSP annihilation cross-section. Typically, this {\it can} be expressed  in a relatively short closed form to reasonable accuracy (albeit in terms of our masses and couplings which would at best be horrendous expressions involving not only the eigenvalues of the mass matrices but the diagonalisation matrices as well), but again there are several regions of parameter space where subtle effects come into play such as so-called co-annihilation, where several states are competing to be the LSP, which  enhances the annihilation rate.  It would at this stage come as no surprise to our omniscient being to discover that these points in the parameter space of the CMSSM are the ones that give the most favourable relic densities. The most interesting points in parameter space are not typical of this particular BSM model  in other words, so one requires not only the LSP cross-sections, but also the cross-sections involving other particles of similar mass (commonly referred to as Next-to LSPs or NLSPs) in the spectrum as well. Finally our omniscient being would need to analytically solve the non-linear evolution of the densities of the LSP and NLSPs in the early universe, inserting these cross-sections in order to find the final freeze-out density of the LSP. If it is not yet abundantly clear, finding analytic expressions for the relic density would amount to being able to solve a whole series of mathematical problems, a solution to any one of which is infeasible.

Thus to cope with the lengthy  chain of analysis the scientific community has instead resorted to automated numerical computation, and developed computational ``packages'' that will accept a given a set of input parameters, perform series of computations such as the one above automatically, and spit out the desired observables.  Machine learning can be used to circumvent the computation chain entirely or to better focus on points of interest (for example in  Refs.~\cite{Caron:2016hib,Hammad:2022wpq,deSouza:2022uhk,Romao:2024gjx,Diaz:2024yfu}). However  machine learning is generally non-interpretable and, for the problems we are interested in, usually not explainable either. In the current context it is worth recalling that {\it in principle}  a suitably trained neural-network that takes in the model parameters and spits out the relic density can be thought of as simply a gigantic nested function that has incorporated by trial-and-error the relations above that are impossible to derive through physical analysis. However the mathematical function underlying a neural-network has a huge degree of redundancy because the functional form is fixed by the network topology, and all that we can adjust are weights and biases,  so not much can be learned from it. 

Non-interpretability in this context (that is the difficulty in understanding the dependence of the observables on the input parameters) is an irritation to physicists. This is because a precise chain of physical computation leads to the precise answer, and if we follow the physical logic we can anticipate {\it certain} rough correlations between input parameters and observables. For example, contributions from new (such as SUSY) particles generally die away as their masses grow due to the so-called decoupling theorem \cite{PhysRevD.11.2856}. Therefore observables such as $(g-2)_\mu$, the anomalous magnetic moment of the muon, generally decrease with increasing values of the three SUSY-breaking parameters  $m_{1/2},~m_0, ~A_0$, and physicists have become rather good at locating regions of parameter space that will yield physically interesting results. This suggests that  some degree of explainability should still be possible that incorporates this and similar physical dependencies.  

Clearly, then, there is tension at play between the simplicity of analytic expressions and their effectiveness as regressors for physical observables. This motivates the use of {\it symbolic regression}~\cite{Koza92} (SR), which in a sense provides an automated way to make this trade-off. SR was discussed in generality in Ref.~\cite{Udrescu:2019mnk}, and for specific applications see Refs.~\cite{Butter:2021rvz,Abel:2022nje,Bartlett:2022kyi,Koksbang:2023sab,Sousa:2023unz,MaurizioEtAl,Bahl:2025jtk,Bendavid:2025urn}. 
 A comprehensive overview SR methods is given in Refs.~\cite{defranca,DBLP:journals/corr/abs-1909-05862,DBLP:journals/corr/abs-2006-11287,DBLP:journals/corr/abs-2107-14351,cranmer2023interpretable}.

What SR aims to provide is a set of analytic expressions for the observables, by learning the analytic formul\ae\/ that can best reproduce their observed values and, like the neural-network, its rationale is merely to fit the data. In fact it is {\it morally} like a neutral-network that is able to change its topology and activation functions in order to more efficiently fit the data. However this potentially gives it the power to yield explainable results in the realm of physics, because if simple correlations such as those alluded to above are important in some region of parameter space then they should manifest themselves in the symbolic expressions it produces. One could for example imagine taking the symbolic expressions for $(g-2)_\mu$ and expanding them around the decoupling limit to find analytic approximations that work in that region of parameter space and that have a physical interpretation. Of course this possibility is contingent on the asymptotic region carrying enough weight in the data provided for the symbolic regression.

The purpose of this paper is to explore the use of symbolic regression in the realm of BSM physics. First we will demonstrate that it is indeed a formidable tool for performing global fits on the CMSSM. To do this we present  a set of analytical expressions for the Higgs mass, $m_{H^0}$, the supersymmetric contribution to the muon anomalous magnetic moment, $\delta(g-2)_\mu$, and the dark matter relic density, $\ODM h^2$. These expressions are functions of the four input parameters of the CMSSM. This part of the study  is based on the work in Ref.~\cite{AbdusSalam:2024obf}. The expressions have been made available at  \cite{symbolic_regression_bsm_2024} alongside the code that produced them:  the dataset used can also be found at~\cite{abdussalam_2024_11366471}. In addition we provide a ``classifier'' function $C(m_{1/2},m_0,A_0,\tan\beta)$ that can decide if a given point in parameter space is likely to be viable based on several physical criteria. 

As an example of these, a particularly troublesome aspect of supersymmetric theories is that, because they introduce many new scalars, they are susceptible to developing instabilities. This happens along particular directions in field space that are relatively flat,  
along which the new scalar superpartners are able to get VEVs themselves, just like the Higgs (which would end in catastrophe). These directions, which are usually referred to as {\it charge and colour breaking (CCB) directions}, must be sufficiently lifted by the choice of parameters to prevent this occurring. Another constraint that can be included in the ``classifier'' function is a fact that the LSP in the spectrum which will become our dark matter particle should of course be uncharged. This implies that it should be composed of uncharged gauginos and higgsinos, and not for example selectrons, and this again requires detailed knowledge of the RG running and subsequent low-energy spectrum. The classifier embodies all of these constraints in one function $C(m_{1/2},m_0,A_0,\tan\beta)$, returning a value greater than $0.5$ if a point is physically viable, and less than  $0.5$ when it is not. To list the criteria we use, they are absence of CCB minima, absence of charged LSP, and absence of negative DM relic density. This information is provided by the packages for each data point, with the point being given a label value of 1 or 0 depending on its viability.

We will demonstrate that it is possible to use the expressions furnished by symbolic regression to perform global fits and  determine the posterior probabilities of the CMSSM input parameters very rapidly. In comparison with the usual methods in which one has to complete the entire chain of computation for each ``live point'' that one may be considering, we will see that fits carried out in this manner can be performed orders of magnitude more quickly. 

We will also explore a  second interesting potential benefit of symbolic regression which is that one may restrict the allowed symbolic expressions to be differentiable. One can do this either by including only differentiable operators in the pool of operators available to the symbolic regression, or by taking previously derived symbolic expressions and replacing non-differentiable operators with differentiable approximations. For example the binary operator max$(a,b)$ (which appears to be quite important in our ultimate expressions at~\cite{symbolic_regression_bsm_2024}), can be replaced everywhere by a 'softmax'  function, for example   
\begin{equation}
\label{eq:softmax}
\max(a, b) ~\approx~ {\rm softmax}(\kappa;\, a, b) ~= ~ \frac{a + b}{2} + \frac{(a-b)}{2} \tanh(\kappa (a - b))~\com
\end{equation}
which approaches the true max function exponentially quickly as $\kappa\to \infty$. 

Having differentiable expressions is advantageous for several reasons. It allows for gradient-based optimisation such as gradient-descent which can for example help in optimising experimental setups. In addition having differentiable expressions from symbolic regression would allow the possibility of  asymptotic analysis (as alluded to above), in which one analyses the behaviour of observables in limits of large/small parameters, possibly regaining explainability in some regions of parameter space. Error propagation calculations become more manageable, as derivatives can be used to estimate how measurement uncertainties in parameters feed through to the observable. Differentiable formul\ae\/ for observables would also allow straightforward calculation of sensitivities (derivatives) with respect to parameters, and help identify which parameters most influence the observable, guiding BSM model building. In the particular context of BSM one particularly important measure of the sensitivity is the so-called {\it fine-tuning}, which in some sense tells us how ``likely'' the model is if we assume uniform prior probabilities of parameters. This has become an important (albeit somewhat controversial) indicator of the  credibility of a BSM model. The most well known of these is the Barbieri-Giudice measure of fine-tuning given in Ref.~\cite{BARBIERI198863}. If we denote the parameters defining the model generically by $p_i$, then the Barbieri-Giudice sensitivity of the Higgs mass-squared  $m^2_{H^0} $ to $p_i$ is given by 
 \begin{equation}
\Delta_{p_i} ~=~   \frac{\partial \, \log {m_{H^0}^2}}{\partial \, {\log p_i}}~.
\end{equation}
The overall measure of fine-tuning (or overall measure of parameter sensitivity) is often taken as the maximum over all  the parameters:
\begin{equation}
\label{eq:BG}
\Delta ~=~ \max_{p_i}  \Delta_{p_i}~.
\end{equation}
Clearly having access to differentiable expressions for the Higgs mass   $m_{H_0}^2 (p_i) $ (and any other observables one may be interested in assessing the fine-tuning for) would be a crucial advantage for such questions.

\section{Symbolically regressing the CMSSM}

\label{sec:symCMSSM}
We now turn to the practicalities of performing SR in BSM studies.   Let us first give a brief overview of the method we developed for our study which would we believe will be applicable to most BSM models. We will then focus on the CMSSM to demonstrate its effectiveness. 

\subsection{Generalities of BSM symbolic regression}

\label{subsec:generalities}

For BSM studies we believe Operon (and the its Python version  PyOperon)~\cite{Burlacu:2020:GECCOcomp}, are optimal due to the efficient vectorised framework and low resource demands. This makes it particularly suitable for BSM studies because they have  multi-dimensional parameter spaces, which in turn necessitates training the symbolic regressor on as much data as possible. In addtition BSM models tends to contain important localised regions of interest containing physical poles and mass-degeneracies, and the SR expressions that are generated need to encompass this highly localised behaviour. Such regions incorporate those physical aspects such as the co-annihilation effect alluded to in the introduction, that can be crucial in finding a good fit for an observable. 

Consequently, successfully training a symbolic regressor on a BSM model will inevitably require  fine multi-dimensional training data.
The data points  obviously have to be pre-run with available packages to generate values of observables which can then be incorporated into  the loss-function of each individual in the population using the symbolic formul\ae\/  generated by its expression trees for the observables in the study.  

Operon is built upon Genetic Programming (GP) which assembles a population of symbolic expressions from a bank of pre-chosen functions, and then evolves it using repeated cycles of {\it selection}, {\it breeding}, and {\it mutation}. Operon incorporates two objectives into the GP loop: one of these is the closeness of the formul\ae\/  to the values generated by the packages, which can be expressed in terms of a number of metrics. 
The regression metric for this can be  ``mean square error'', ``mean average error'', ``$r^2$'', or ``normalised mean square error''.
The other objective for the GP loop is the expressions ``length'', which is simply the string length of the mathematical expression. By evaluating the individuals in the population on these two metrics, Operon is able to rank them, and select them for breeding at the start of each cycle. It then performs breeding and mutation to generate a new population, and repeats the process to allow the population to evolve until it contains good symbolic expressions which lie along the so-called  {\it Pareto front}. The final output of the Operon run is the population of individuals inhabiting the  Pareto front.

Operon has a number of additional hyperparameters, for example the population size and the population initialisation (see Ref.~\cite{Burlacu:2020:GECCOcomp}).
To optimise these hyperparameters as well as the regression metrics, we use Optuna loops~\cite{optuna_2019}. 
As the ``figure of merit'' for the choice of hyperparameters, one can use the relative error of the observables for the Pareto front individuals.
This is preferable to other common regression metrics, such as $r^2$, because the latter can be highly biased by large observable values, when the observable varies by orders of magnitude within the parameter space. 

Let us briefly return to the question of those observables such as the dark matter relic density which may be hard to fit within a particular BSM model. As described above this can arise due to the viable values of the observable requiring some physical phenomenon to take place which turns out to occur only in certain restricted region of parameter space. In such a situation one can find that the symbolic expressions are driven by the `wrong' regions of parameter space in an attempt to  fit the bulk of the data, but are not accurate in the regions of most interest.  For an observable such as the dark matter relic density the problem is exacerbated by the fact that the generic value can be orders of magnitude larger than desired. In such cases, there are ever diminishing returns from larger budgets and maximal tree-sizes. This can  be addressed by weighting the data so that it better favours the regions of interest for those observables. In particular we would like the expressions produced to be able to capture the more subtle behaviour that leads to the desired cancellations. In order to do this one can operate by resampling the training data according to ``flattening weights''. These level the distribution to ensure that the regressor is not overly dominated by fitting the observable around its most common values, as opposed to the values of interest. Indeed the weights can be adjusted to resample around the areas of most interest.

For difficult to fit observables like the  $\ODM h^2$, without resampling the training data in this manner, SR is unlikely to accurately map the physical region of interest (i.e. $\ODM h^2\sim \mathcal{O}(0.1)$ for the relic density).
However, as we will see there can still be considerable contamination in the range of physical interest (i.e. points that the symbolic expressions indicate have reasonable dark matter relic density when actually they do not, and vice versa). This kind of problem is not present for observables that have acceptable {\it generic} behaviour within a particular BSM model, as for example the Higgs mass does within the CMSSM. In this sense the latter can be considered to be an ``SR-easy'' observable within the CMSSM while  an observable like $\ODM h^2$ that requires re-weighting is an ``SR-hard'' one. The labels ``easy'' and ``hard'' are in this context obviously entangled with our own preference for what the observable should be, and  crucially are BSM model dependent. Conversely, symbolic regression gives us a crude working criterion for determining how {\it natural} a particular BSM model is. Namely, the most natural BSM model with the least ``fine-tuning'' in the sense of Eq.~\eqref{eq:BG}, is likely to be one in which all the observables are found to be SR-easy.

\vspace{-0.2cm} 
\subsection{Analytic expressions for the CMSSM}

We now turn to the specific details of the CMSSM. As discussed the model is defined by  four parameters  $m_{1/2},~m_0, ~A_0$ and $\tan\beta $ (in addition to the couplings and masses of the Standard Model upon which it is built). 
Our data was composed of $10^5$ points sampled at random from the parameter space. We take uniform prior probabilities in the intervals $[0, 10]$~TeV for $m_{1/2}$ and $m_0$, $[-6, 6]$~TeV for $A_0$, and $[1.5, 50]$ for $\tan\beta$. The data was split into a training and a test set, and the observables  $m_{H^0}$, $\ODM h^2$, and $\delta (g-2)_\mu$ were determined for each data point using the particle spectrum generator SPheno~\cite{Porod:2003um} and the particle dark matter package MicrOMEGAs~\cite{Belanger:2001fz}. In addition the physical viability of each point was determined (and given a value $1$ or $0$) which feeds into the classifier regressor. 

PyOperon was run on this sample of points using the re-weighting procedure for $\ODM h^2$ described in subsection \ref{sec:symCMSSM}\ref{subsec:generalities}, which essentially flattens the weighting such that all values of $\ODM h^2$ carry equal weight within the regression. We found that the re-weighting reduced the effective training dataset from $10^5$ to $10^4$ data points, and gave an approximately flat distribution of $\ODM h^2$. (As we shall see this will improve the ability of the symbolic regressors to fit $\ODM h^2$ to its inferred value.) The result is the set of  symbolic regressors for $m_{H^0}$, $\ODM h^2$, and $\delta (g-2)_\mu$ and the classifier $C$ which have been made available at  \cite{symbolic_regression_bsm_2024}.

\begin{figure*}[t]
  \centering
  \vspace{-1cm} 
  \includegraphics[scale=0.28]{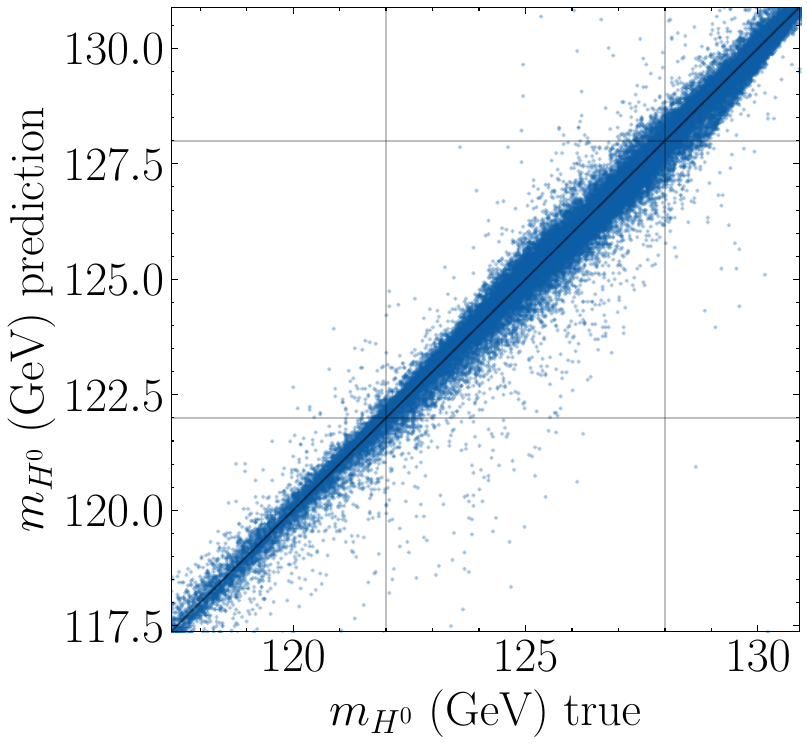}
  ~~~~~~\includegraphics[scale=0.28]{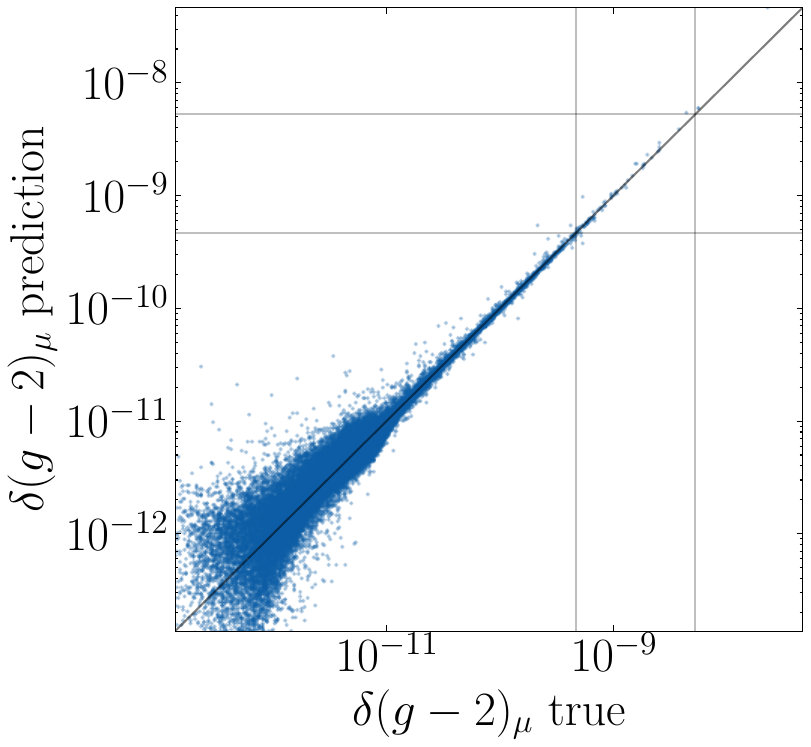}
  ~~~~~~\includegraphics[scale=0.28]{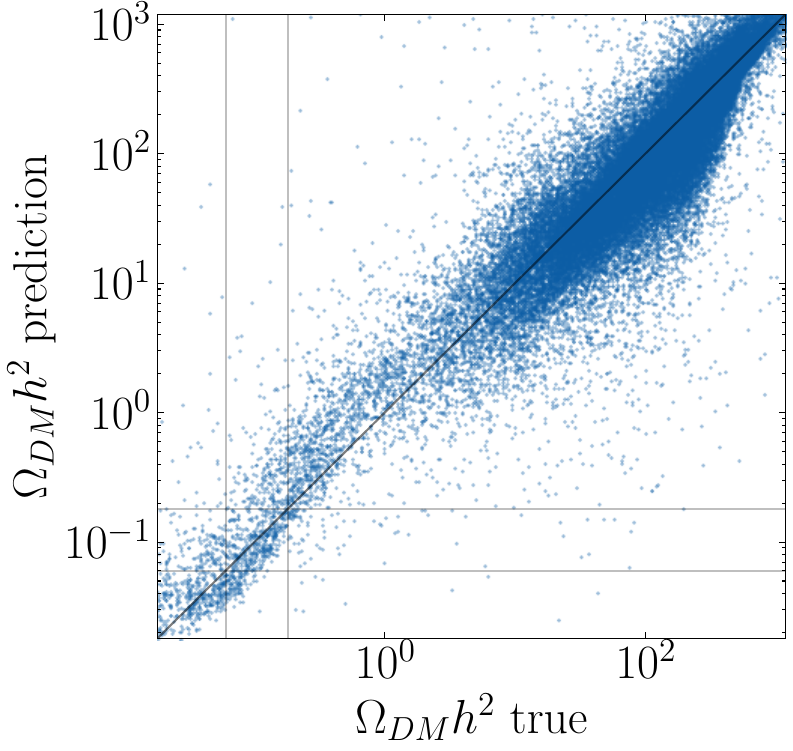}
  \includegraphics[scale=0.28]{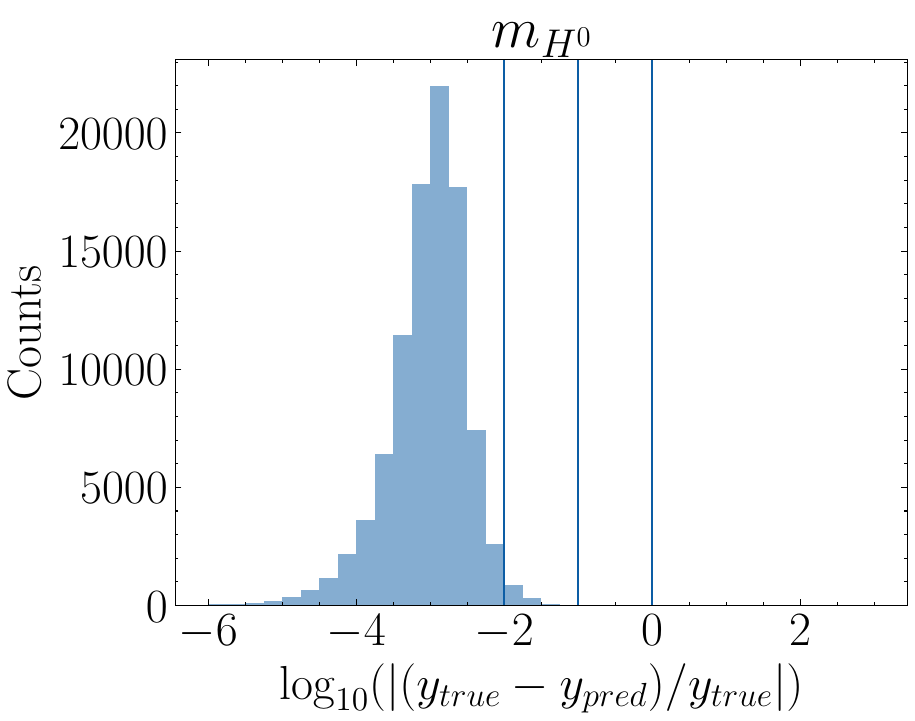}
  \includegraphics[scale=0.28]{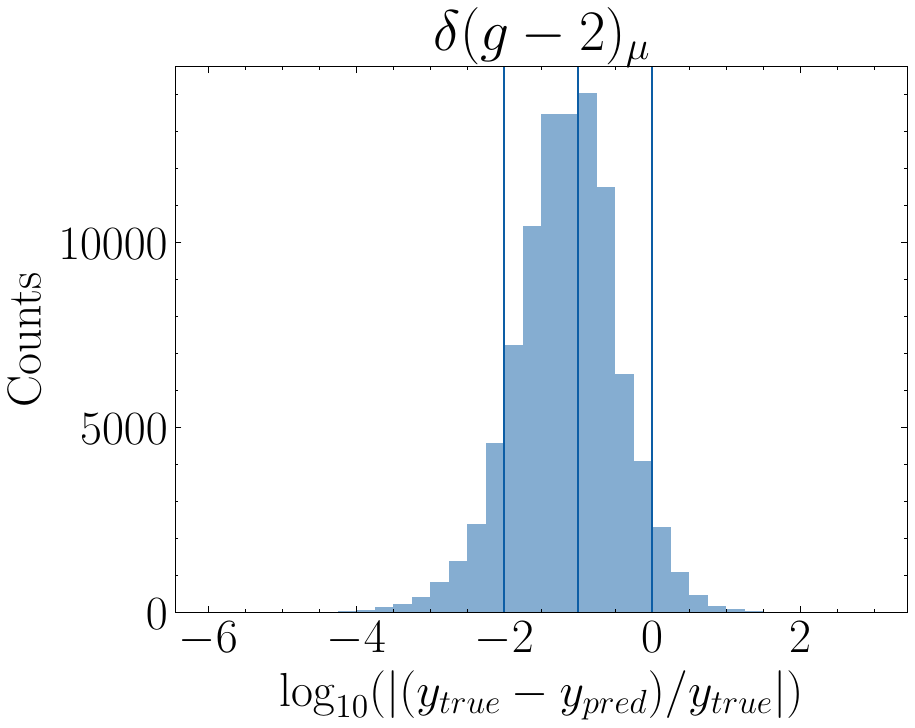}
  \includegraphics[scale=0.28]{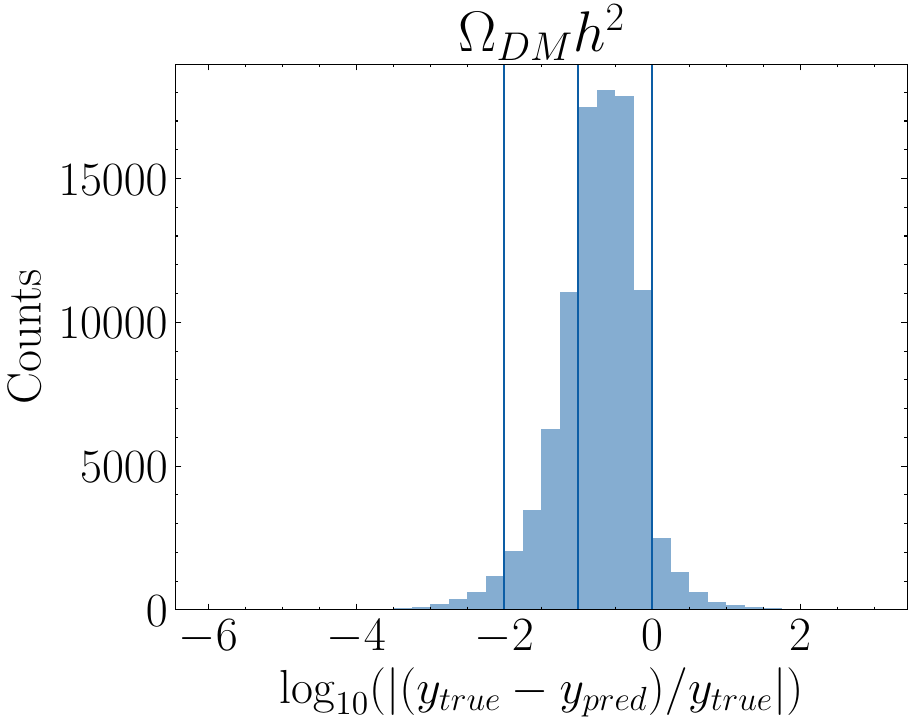}
  \caption{The performance of the SR expressions for the observables, on the test set. In the upper panels we show the ``true vs. prediction'' scatter plots. The solid lines delineate the physically viable values. In the lower panels we show the relative errors. Here the vertical lines delineate relative errors of 1\%, 10\%, and 100\%.}
  \label{fig:true-pred-and-rel-errors}
\end{figure*}

\begin{figure*}[t]
  \centering
  \includegraphics[scale=0.4]{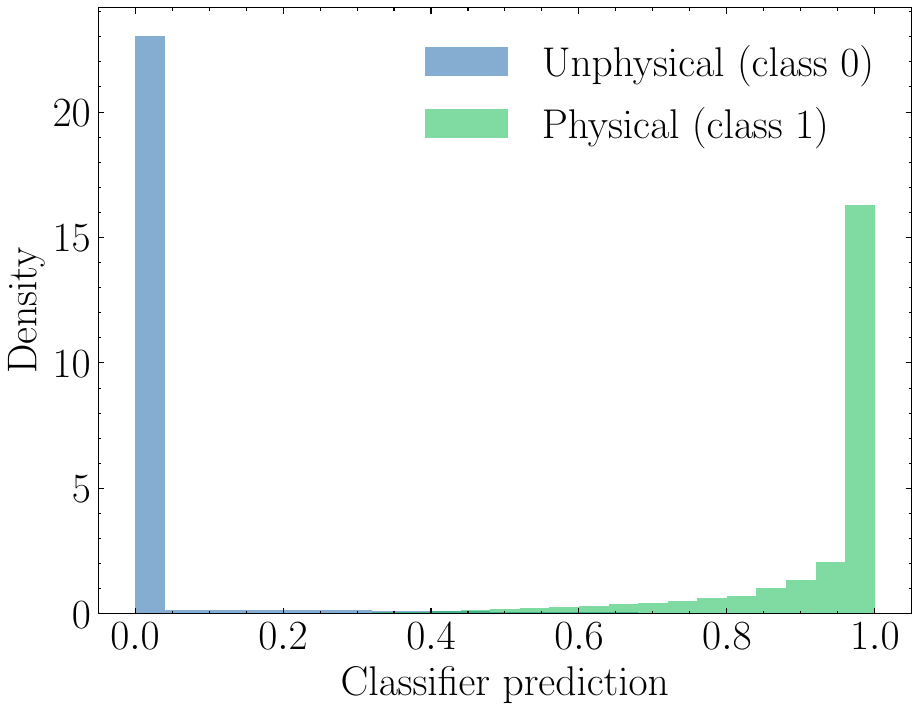}
  ~~~~~~\includegraphics[scale=0.4]{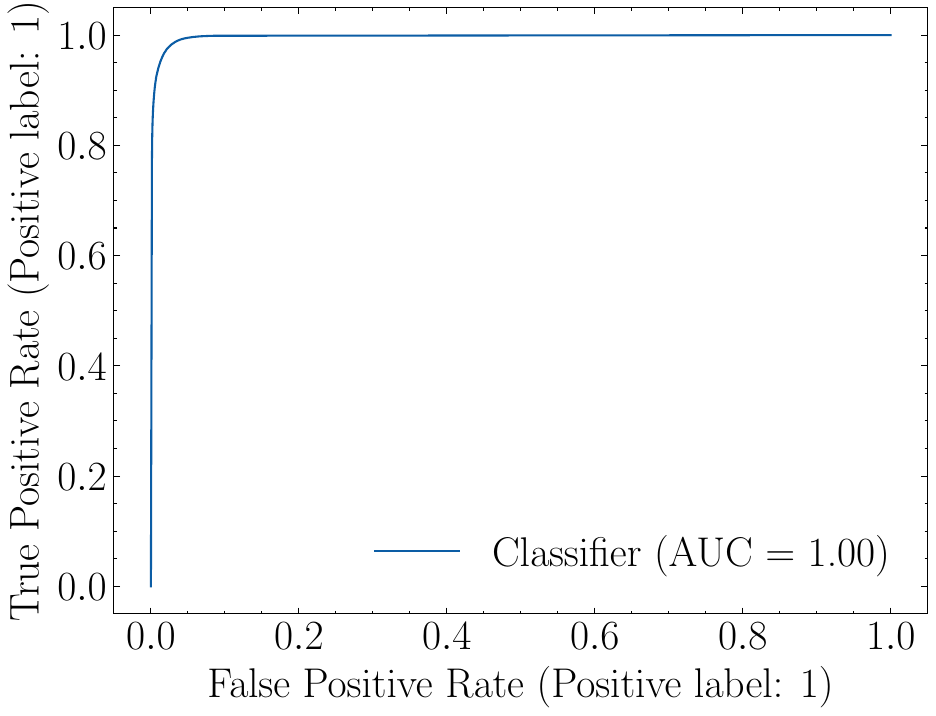}
  \caption{{The classifier symbolic regressor performance, showing the output of the classifier when acting on the test data on the left and the ROC curve on the right.  } }
  \label{fig:classifier}
\end{figure*}

To test the expressions we present in the upper panels of ~\cref{fig:true-pred-and-rel-errors} scatter plots (made using the test set) of the ``true'' value of the observable (denoted generically $y_{\rm true}$)  versus the prediction (denoted $y_{\rm pred}$) from the symbolic regressor.  The relative errors of the predictions, defined as  
$
|y_{\rm pred} - y_{\rm true}|/|y_{\rm true}|\com
$
are presented in the lower panels. The regression of the Higgs mass is evidently very accurate, with almost all  points having relative error below 1\%. For $\delta (g-2)_\mu$ the errors are also extremely small in the region of physical interest where $\delta(g-2)_\mu$ is enhanced, and actually smaller than the theoretical value of $(g-2)_\mu$ in the Standard Model itself. For $\ODM h^2$ the {\it relative} errors are typically  above 10\%, but this is actually better than the 20\% theoretical error that can be assigned to $\ODM h^2$. Finally the performance of the classifier regressor is shown in the two panels of Fig.~\ref{fig:classifier}, as a distribution and as a ROC curve. From these plots we see that (by accepting only points with say $C>0.5$) we have an almost perfect classifier regressor. In other words our function $C(m_{1/2},m_0,A_0,\tan\beta)$ can be used to immediately reject unphysical points from the CMSSM.

It is worth commenting on the question of whether the symbolic regression has converged, or if better expressions could be found. To consider this question, we produced a ten times larger data set and generated a new set of symbolic expressions. It was found that these new expressions did not demonstrate any significant improvement in accuracy, while the method became prohibitively slow. It is worth noting that the scaling did not appear to be linear with data set size.

\section{Global fits}

In order to put our symbolic expressions to the test let us now consider performing a global fit, that is a statistical analysis to determine the most probable values of the CMSSM model parameters. In this work we shall use  
Dynesty~\cite{2020MNRAS.493.3132S}. This package performs nested sampling, a Bayesian technique used for parameter estimation and calculating the model evidence. It is a very efficient method of exploring complex, multi-dimensional parameter spaces, making it well-suited for BSM global fits. The process begins by defining a likelihood function that combines experimental and theoretical constraints into a single metric indicating how well a set of parameters fits the data. Using our previously specified flat priors for the parameters, Dynesty initialises a set of ``live points'' sampled from these priors and iteratively replaces the lowest-likelihood point with new samples drawn from the prior constrained to higher likelihood regions. This iterative process systematically explores the parameter space, effectively mapping out the posterior distributions and estimating the Bayesian evidence.

This procedure can come with significant resource overheads due to the need in each iteration to find a new point with a likelihood greater than that of the lowest likelihood point in the set of live points.  Hence, we will now show that replacing these costly full physics estimates of observables with symbolic expressions dramatically speeds up the process with only a modest loss of accuracy.  

For these fits we use $m_{H^0} = 125.04 \pm 0.12$~GeV~\cite{CMS-PAS-HIG-21-019} and $\ODM h^2 = 0.12 \pm 0.02$~\cite{Planck:2018vyg} which includes an aforementioned theoretical uncertainty of 20\% in the relic density. The experimental value of the anomalous magnetic moment of the muon is $\delta (g-2)_\mu = (249 \pm 48) \times 10^{-11}$~\cite{Muong-2:2023cdq}. This represents a discrepancy with its high precision Standard Model (SM) prediction which is of current interest~\cite{Aoyama:2020ynm}. 
Denoting the data for these three observables generically as $\underline{d} = \{ \mu_i \pm \sigma_i \}$, where $\mu_i$ and $\sigma_i$ represent the central values and uncertainties (with $i=1,2,3$), then for each point $\underline{\theta} = \{ m_{1/2}, m_0, A_0, \tan \beta\}$ we shall estimate the likelihood as
\beq 
p(\underline{d} | \underline{\theta}) ~=~ \prod_{j=1}^3 \, \frac{1}{\sigma_j \sqrt{2\pi}} \, \exp\bigg\{ \frac{-(y_{\rm pred}^j - \mu_j)^2}{2\sigma_j^2} \bigg\}~.
\eeq 
Here $y_{\rm pred}^j$ stands for  the predictions given by the SR expressions for the  observables. For comparison with the nested sampling results, version 2.1.4 of Dynesty was employed along with \texttt{DynamicNestedSampler} with 500 live points. (For the other parameters, we took no \texttt{bootstrap}, with \texttt{pfrac=1}.)

\begin{figure*}
  \centering
\includegraphics[width=0.8\textwidth]{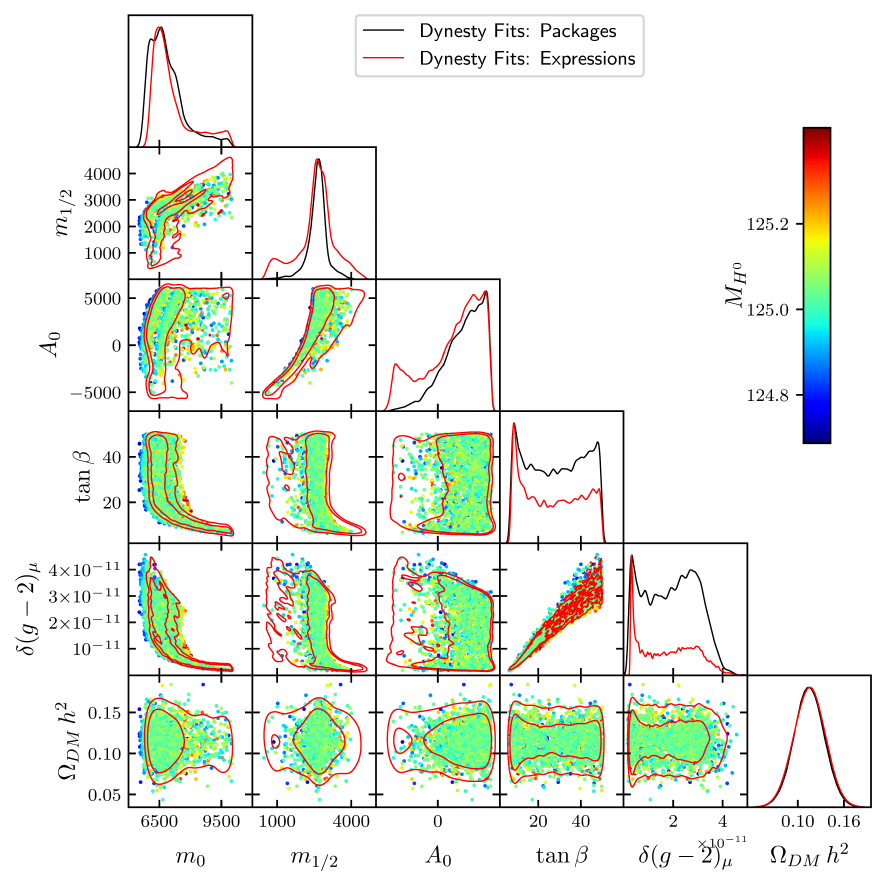}
  \caption{{Posterior probability distributions for the CMSSM found using Dynesty, fitting the three observables  $m_{H^0}$, $\delta (g-2)_\mu$, and $\ODM h^2$. The package-based results are the scatter plots (and the black lines on the diagonal plots). The SR  expression-based results are the red lines. Dimensionful parameters are in GeV. As is customary the 1D marginals are normalized with respect to their maximum heights rather than areas in order to aseess the agreement of maximum a posteriori probabilities.}}
  \label{fig:tri_mh_dyn}
\end{figure*}

In \cref{fig:tri_mh_dyn} we show the resulting posterior distributions of the CMSSM parameters, which have been fit to our three observables using Dynesty, (with the plots produced with GetDist~\cite{Lewis:2019xzd}). 
The 68\%  and 95\% Bayesian probability contour lines (in red and labelled ``Expressions'') are the posterior distributions from the fit which were determined using the SR expressions. 
Meanwhile the scatter plots (with colour denoting the Higgs mass, and labelled ``Packages'') show the global fit made using conventional methods, in which our three observables were computed using the available packages. 

We can conclude that the global fit analysis made using SR derived expressions compares extremely well with the fit made using the conventional packages-based method. However we should comment on the apparent differences in the 1D marginal plots. The most important point to make is that the plots are normalized by maximum height rather than area. This is the customary approach in order to compare the locations of maximum a posteriori probabilities. This makes the $\tan \beta$ 1D marginal appear to be in worse agreement than it really is. (In area normalized plots the two graphs would overlap over most $\tan\beta$ values with a somewhat larger spike appearing at low $\tan\beta$ in the plot made using expressions.) A second point is that in Ref.~\cite{AbdusSalam:2024obf} a comparison was made between the fits made using packages with different sampling frameworks to each other. Namely we compared fits made using Dynesty and MultiNest with customary packages, and found that the fits made using these alternative frameworks showed similar minor discrepancies amongst themselves. Therefore we consider the results found using expressions to be {\it as reliable} in this sense as the results found using packages.

\section{Differentiable Symbolic Regression}

We now turn to the question of utilising entirely differentiable expressions for symbolic regression. 
The most straightforward approach for this is to simply take the previously used approximations and replace their non-differentiable parts by differentiable approximations such as the softmax function in Eq.~\eqref{eq:softmax}.
Indeed for our application, the only non-differentiable operators of concern are $\min$ and $\max$, which only appear in the classifier. 
Since we do not use this in the MCMC, we already have functions comprised of differentiable operators, and thus write these in \textsc{jax} so that we can utilise its auto-differentiation. Of course, given that these are symbolic expressions, one could also analytically compute these derivatives.

Now that we have access to differentiable surrogate models for the ``packages,'' we can leverage differentiable optimisation or sampling techniques to expedite our analysis. To demonstrate this, we replace the nested sampler Dynesty with a No-U-Turn Sampler (NUTS) \cite{Hoffman_2011}, as implemented in \textsc{numpyro} \cite{Bingham_2018,Phan_2019}.
Using the same likelihood as before, we run 4 parallel chains with 5000 warm-up steps and $10^5$ samples each.
We verify that this results in Gelman-Rubin statistics \cite{Gelman_1992} which are unity within $10^{-2}$ for each inferred parameter.

We plot the inferred posteriors for our expressions in \cref{fig:tri_nuts}, where we compare against the posterior obtained using Dynesty. One sees that we obtain similar agreement between the posteriors as in \cref{fig:tri_mh_dyn}, 
with the same locations of the maxima, and reasonably similar marginalised one-dimensional posteriors. The match is not perfect, however, but it was also found by Ref.~\cite{AbdusSalam:2024obf} (from where we take the Dynesty contours) that the posteriors obtained by Dynesty and Multinest (an alternative nested sampling algorithm) are also not identical. 

Interestingly, we find that the expressions given in Ref.~\cite{symbolic_regression_bsm_2024} appear to have some instabilities when we attempt to use their gradients in our MCMC. This manifests itself as a large number of divergences, which indicates a numerical instability during Hamiltonian trajectory integration. This is typically caused by regions of high curvature or poorly scaled parameters in the posterior distribution. We circumvented this problem by increasing the target acceptance probability of NUTS to be 0.95 instead of the default 0.8. This increases the probability of rejecting bad steps and thus fewer divergences are encountered. This essentially sacrifices efficiency for stability.

Upon more careful investigation, we identified the function approximating $\Omega_{\rm DM}h^2$ as the cause of our divergences. We therefore generated a new symbolic regressor for this variable, which only contains differentiable operators. This was obtained by restricting PyOperon allowed operators to their differentiable subset, i.e. operators such as $\min$ and $\max$ were disallowed during the Optuna optimisation loop.

Replacing the original expression with this expression enables us to keep the target acceptance probability at 0.8 without encountering the same number of divergences. We find sampling to be much more efficient; the run time on a MacBook Pro with an Apple M2 Pro chip (12-core CPU) and 16 GB of unified memory for the original expressions with NUTS is approximately 25 minutes, but for the same number of samples we only require 4.5 minutes for the new expression.
(This could be accelerated further by utilising the GPU, but we only used a CPU in these experiments.)
'This can be compared with the time required to sample this posterior with Dynesty 
which took about 2 hours on 72 cores using the packages, or 2 hours on a single core using the expressions\footnote{For the expressions, the overhead of parallelising the run outweighs the benefits as expressions are computationally cheap to evaluate.}, and thus we see a substantial improvement for either expression. 

\subsection{Comparison with Neural Networks}

For completeness, it is worth comparing at this point the performance of the SR derived expressions to the performance of a neural network (NN), as these are also differentiable, and would also allow for a differentiable fit. To this end, we adopted a
fully connected feed-forward neural network in \texttt{JAX}
to produce differentiable likelihoods, so that posterior sampling could be performed in the same manner (NUTS) as for the differentiable symbolic expressions. 
The model consists of four hidden layers with 64 neurons each, mapping a 4-dimensional input to a 3-dimensional output. Parameters were initialised with a fixed random seed for reproducibility and trained using the Adam optimiser \cite{Kingma2014} with an initial learning rate of $10^{-3}$.
Given the different dynamic range and observational uncertainties of our outputs, instead of simply using a mean-squared-error (MSE) loss function, we weight each term in the MSE loss by the squared reciprocal of the observational error on that quantity.
The learning rate was reduced by a factor of 0.5 if the validation loss did not improve for 200 epochs, with a minimum of $10^{-6}$, and training was performed for up to 30,000 epochs.

We can make an interesting immediate observation about the NN performance. Namely, under the original priors and with the original training samples the NN failed to capture the same global picture that the SR expressions were able to provide out of the box. Therefore, in order to obtain meaningful posteriors for the NN regressors, substantially more data had to be generated in order to create a more tightly focussed training domain than was required for the SR to perform.  To take account of this, the NN was trained using a much larger dataset ($10$ million points), split $90\%:10\%$ for training/validation, with a restricted prior window for the input parameters of
\begin{align}
    m_0 & : [3.5,10]\ \text{TeV} \\
    m_{1/2} & : [0.02,6]\ \text{TeV} \\
    A_0 & : [6,6]\ \text{TeV} \\
    \tan\beta &: [5,50]~.
\end{align}
In addition, a ``cut'' was enforced on the data to allow only those points that produced observables in promising ranges, namely 
\begin{align}
    \Omega_{DM}h^2 & \in [0,0.3] \\
    m_{H_0} & \in [122,128]\ \text{GeV} \\
    \delta(g-2)_{\mu} & \leq 10^{-10} \ ,
\end{align}
producing a dataset ``zoomed in''  on the region of interest.

For a fair comparison, we then retrained the SR expressions on the same enlarged and ``zoomed'' dataset that was used for the NN. 
One substantial difference in behaviour emerged which was the greater sensitivity of the SR to those points which then lay outside the domain of the training data which produces promising observables. To address this  we trained a classifier to discard points lying outside the ``promising observables'' window. This is operationally equivalent to enforcing a sharper prior and is consistent with the way that low-likelihood points would be treated during sampling.
Interestingly, the NN showed much less sensitivity to these ``undesirable'' regions, providing good fits regardless of the use of such a classifier. 

\begin{figure*}
  \centering
\includegraphics[width=0.8\textwidth]{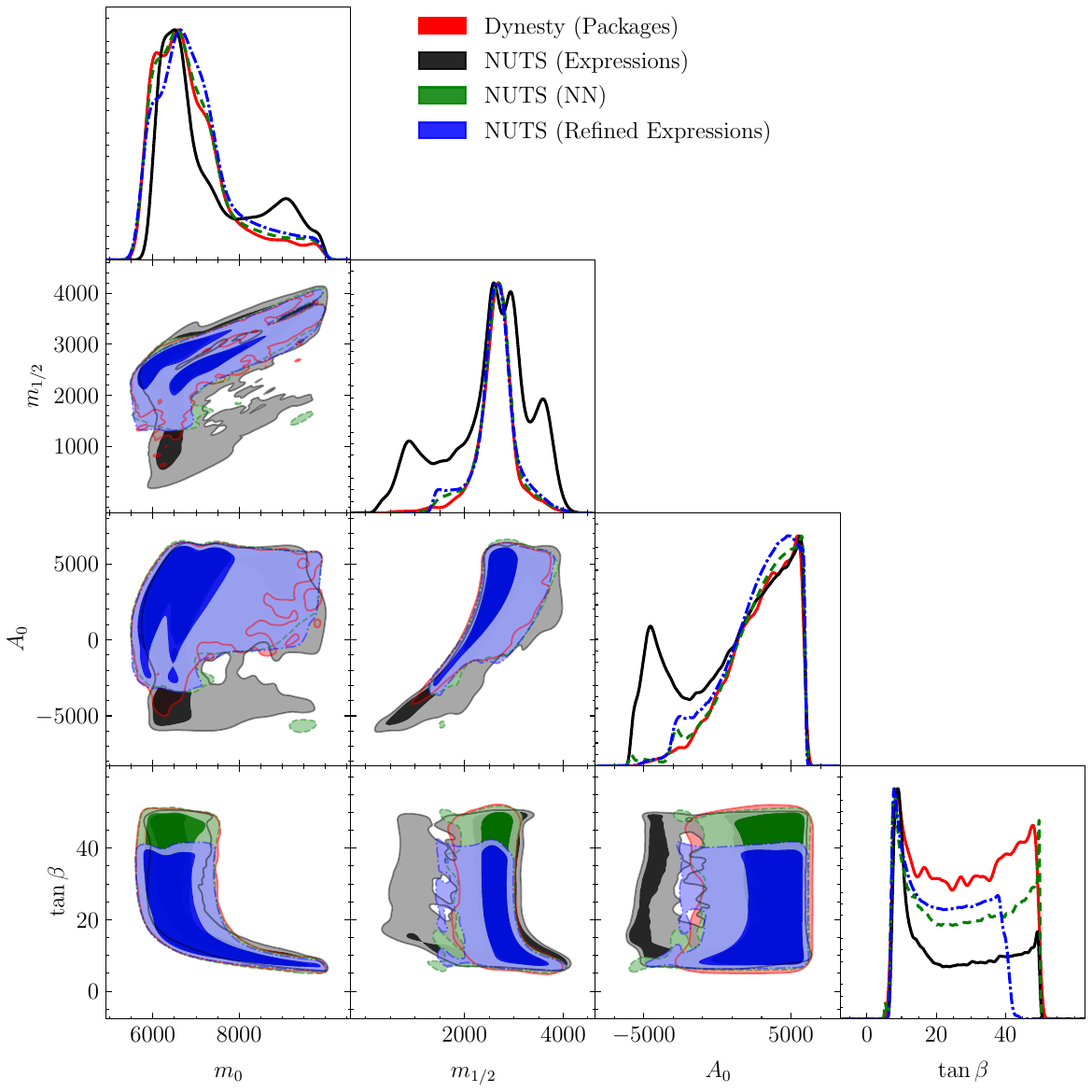}
  \caption{Posterior distribution of CMSSM parameters obtained using NUTS compared to those from Dynesty. The contours show the 68\% and 95\% confidence intervals. 
The red contours compute the outputs exactly, whereas the remaining ones use the expressions of Ref.~\cite{AbdusSalam:2024obf} (black), a neural network (green), or refined symbolic expressions for a narrower prior range (blue) to bypass this calculation. The 1D marginals are normalized with respect to their maximum heights rather than areas in order to assess the agreement of maximum a posteriori probabilities.  }
  \label{fig:tri_nuts}
\end{figure*}

Under these matched conditions the NN and SR posteriors became very similar and both agreed with the package-derived posterior (see Fig.~\ref{fig:tri_nuts}, where the new SR result is shown in blue contours). 
For the three well-constrained parameters ($m_0$, $m_{1/2}$, and $A_0$), we see that the refined SR contours match well those from either Dynesty or the NN, and the marginal posterior for $m_{1/2}$ is much more accurate than when obtained using the regressors of Ref.~\cite{AbdusSalam:2024obf}. For both cases, the marginal distribution of $\tan\beta$ does not match perfectly, and for the refined SR expressions this becomes zero at slightly smaller values than for the truth.
Hence, both the NN and refined SR expressions achieve good approximate posteriors, albeit with some small remaining artefacts.

The main differences in the performance of the NN versus the SR are therefore mainly practical: it seems that a NN approach requires a focused prior dataset in order to reach the same fidelity. By contrast, SR reaches comparable fidelity with substantially less focussing of the prior dataset and shows better global behaviour in this context.

\section{Conclusion} 

Analytic expressions for observables are a powerful tool, however although some observables in a BSM model can be  relatively easily treated with a full physical analysis, the vast majority of them  cannot. In supersymmetric extensions of the Standard Model this is true even at the level of the mass spectrum (which in supersymmetry entails the diagonalisation of $6\times 6$ matrices), before it is even fed into observables such as the dark matter relic density. It is worth noting that the Higgs mass itself receives crucial one-loop and two-loop corrections. 

In the absence of general analytical formul\ae, the typical approach for BSM physics is to perform a computation of the values of observables for any given value of the model's input parameters, using available numerical packages. This results in a costly chain of computation that makes it difficult to effectively analyse the phenomenology of a BSM model such as the CMSSM over the whole parameter space, even using nested sampling techniques. There are indeed cases where it may take several minutes to compute the observable for a {\it single point} by conventional methods~\cite{Diaz:2024yfu}. In this work, we have shown that symbolic regression can bridge this gap and provide analytical formul\ae\/  where a fully  analytical treatment of the physics would be too difficult. 

We showed that the symbolic expressions that are produced can successfully short-circuit the chain of physical computation for a given BSM model (specifically the CMSSM for the current study) between the input parameters and the observables. This greatly reduces the computational resources that are needed for  global fitting. Indeed our study found that two orders of magnitude less CPU time was needed to perform a global fit of the CMSSM when all three of the  observables (namely the Higgs mass, dark matter relic density, and anomalous magnetic moment of the muon) were simply evaluated using the SR derived symbolic expressions. As we saw in Fig.~\ref{fig:classifier}, symbolic regression could also produce a remarkably effective symbolic classifier for implementing model constraints (such as absence of charge and colour breaking minima, and neutral dark matter in supersymmetric extensions of the Standard Model).  This allows much more rapid rejection of unviable models. Moreover, in an era when there are still many competing BSM models, symbolic expressions provide an especially useful tool for model comparison, rather than having to correctly implement packages for each model independently.

It is worth commenting on fact that although  the inference with SR expressions is much faster, this does not take into account the fact that there is obviously computational cost in obtaining the training samples. To compare the computational cost of the global fits using expressions versus nested sampling, the number of times that, for example, the dark matter relic density has to be calculated is roughly the same, i.e. approximately a million times in both cases. An important point, however, is that once the symbolic expressions have been generated they can be reused when for example there are new experimental results that warrant new posterior calculations. This therefore amortizes the costs of data generation and expression training.

A second great advantage of symbolic regression is the possibility to produce differentiable expressions for observables, allowing differentiable workflows for model analysis. Operationally there is a preference for the following procedure for differentiable symbolic regression: first perform unconstrained symbolic regression on the observables; then replace non-differentiable operators (like `max') with differentiable approximations (like the `softmax' function of Eq.~\eqref{eq:softmax}). The reason for preferring this procedure (as opposed to performing a symbolic regression with only differentiable operators) can be easily understood if one considers why non-differentiable operators are likely to turn up in the symbolic expressions of, for example, the classifier function. One objective of this function is to reject theories with charge- and colour-breaking minima, and this criterion is typically expressed as an inequality involving combinations of input parameters, which can be expressed as a max function in the classifier. Likewise satisfying experimental exclusion bounds would most likely result in max functions involving the masses, and so forth. In other words, during the analysis of a typical BSM model there are many occasions when the spectrum must be compared with physical values, and this can be implemented efficiently with non-differentiable operators. If one were to  perform the symbolic regression excluding non-differentiable operators in the pool of available operators, then the regressor would be forced to try and rebuild operators such as `max' with approximations such as the `softmax' function of Eq.~\eqref{eq:softmax}. But these approximations would greatly increase the length of the corresponding expressions, disfavouring such expressions in the symbolic regression. We believe this can lead to  poor performance of an entirely differentiable symbolic regression procedure compared to a procedure in which non-differentiable operators are allowed, and simply replaced with approximations in the ultimate expressions. On the other hand, we also found instances, such as for $\ODM h^2$, where it is advantageous to generate symbolic expressions using exclusively differentiable operators from the start. This appears to greatly reduce instability and increase efficiency during the differentiable optimisation when producing global fits. 

Finally, we found that the SR method compares favourably with a neural network approach. While a NN regressor is an appropriate and viable approach for this  task, it produces compatible posteriors only when significant cuts are made on the prior window for the input parameters in order to ``zoom in'' on regions that produce good observables. Symbolic regression, by contrast, proves to be more sample-efficient on this problem (in the sense that it does not require such prior cuts on the data), thus providing a robust global picture over the whole range of observables. The two approaches are therefore somewhat complementary: NNs are flexible and powerful when there is abundant training data that can be then focussed, while SR offers analytic expressions that provide a better global picture, and which can be straightforwardly integrated into differentiable pipelines.\\

\vspace{0.2cm} 
\noindent 
We are extremely grateful to Bogdan Burlecu and Gabriel Kronberger for extensive guidance with Operon. We would like to thank Miles Cranmer and Pedro Ferreira for help and discussions. SA$_{1}$ and SA$_{2}$ thank CERN-TH and SA$_{1}$ thanks the Institute for Theoretical Physics at Heidelberg University for hospitality extended during the initial stages of this work. SA$_2$ and MCR are supported by the STFC under Grant No. ST/T001011/1. 
DB was supported by the Simons Collaboration on ``Learning the Universe'' and is supported by Schmidt Sciences through The Eric and Wendy Schmidt AI in Science Fellowship.
This work was performed using resources provided by the Cambridge CSD3, provided by Dell EMC and Intel using Tier-2 funding from EPSRC grant EP/T022159/1, and DiRAC funding from the STFC.
We thank Jonathan Patterson for smoothly running the Glamdring Cluster hosted by the University of Oxford, where some of the data processing was performed.

\vspace{-0.4cm}
\bibliographystyle{unsrt}

\bibliography{references} 

\end{document}